\documentclass[]{interact}

\usepackage{epstopdf}
\usepackage[caption=false]{subfig}

\usepackage[numbers,sort&compress]{natbib}
\bibpunct[, ]{[}{]}{,}{n}{,}{,}
\renewcommand\bibfont{\fontsize{10}{12}\selectfont}
\makeatletter
\def\NAT@def@citea{\def\@citea{\NAT@separator}}
\makeatother

\usepackage{color}
\usepackage{booktabs}
\usepackage{hyperref}
\theoremstyle{plain}

\theoremstyle{definition}

\theoremstyle{remark}

\usepackage{tabularx}
\newcolumntype{n}{>{\hsize=1.1\hsize}X}
\newcolumntype{w}{>{\hsize=0.9\hsize}X}
\usepackage{pifont}
\usepackage[table]{xcolor}
\definecolor{ForestGreen}{RGB}{34,139,34}
\newcommand{\rot}{\rotatebox{90}}
\newcommand{\T}{\color{ForestGreen}\ding{51}}
\newcommand{\X}{\color{red}\ding{53}}
\newcommand{\Q}{\color{gray!30}$\bigcirc$}

\usepackage{multirow}
\usepackage{makecell}

\begin{document}


\title{Exploring technologies to better link physical evidence and digital information for disaster victim identification}

\author{
    \name{
        David~Lovell\textsuperscript{a,b},
        Kellie~Vella\textsuperscript{a,b},
        Diego~Mu\~noz\textsuperscript{c},
        Matt~McKague\textsuperscript{a},
        Margot~Brereton\textsuperscript{a,b},
        and Peter~Ellis\textsuperscript{d}\thanks{CONTACT David~Lovell. Email: David.Lovell@qut.edu.au}
        }
    \affil{
        \textsuperscript{a}Queensland University of Technology School of Computer Science, Brisbane, Australia;
        \textsuperscript{b}Queensland University of Technology Centre for Data Science, Brisbane, Australia;\\
        \textsuperscript{c}Centre for Design Innovation, Swinburne University of Technology, Melbourne, Australia;\\
        \textsuperscript{d}Griffith University School of Environment and Science, Brisbane, Australia.
    }
}

\maketitle

\begin{abstract}
Disaster victim identification (DVI) entails a protracted process of evidence collection and data matching to reconcile physical remains with victim identity. Technology is critical to DVI by enabling the linkage of physical evidence to information. However, labelling physical remains and collecting data at the scene are dominated by low-technology paper-based practices. We ask, how can technology help us tag and track the victims of disaster?  Our response to this question has two parts. 

First, we conducted a human-computer interaction led investigation into the systematic factors impacting DVI tagging and tracking processes. Through interviews with Australian DVI practitioners, we explored how technologies to improve linkage might fit with prevailing work practices and preferences; practical and social considerations; and existing systems and processes.  We focused on tagging and tracking activities throughout the DVI process. Using insights from these interviews and relevant literature, we identified four critical themes: protocols and training; stress and stressors; the plurality of information capture and management systems; and practicalities and constraints.  

Second, these findings were iteratively discussed by the authors, who have combined expertise across electronics, data science, cybersecurity, human-computer interaction and forensic pathology. We applied the themes identified in the first part of the investigation to critically review technologies that could support DVI practitioners by enhancing DVI processes that link physical evidence to information. This resulted in an overview of candidate technologies matched with consideration of their key attributes. 

This study recognises the importance of considering human factors that can affect technology adoption into existing practices. Consequently, we provide a searchable table (as Supplementary Information) that relates technologies to the key considerations and attributes relevant to DVI practice, for the reader to apply to their own context. While this research directly contributes to DVI, it also has applications to other domains in which a physical/digital linkage is required, and particularly within high stress environments with little room for error.
\end{abstract}


\subsection*{Graphical abstract}
\begin{center}
    \includegraphics[width=11.5cm]{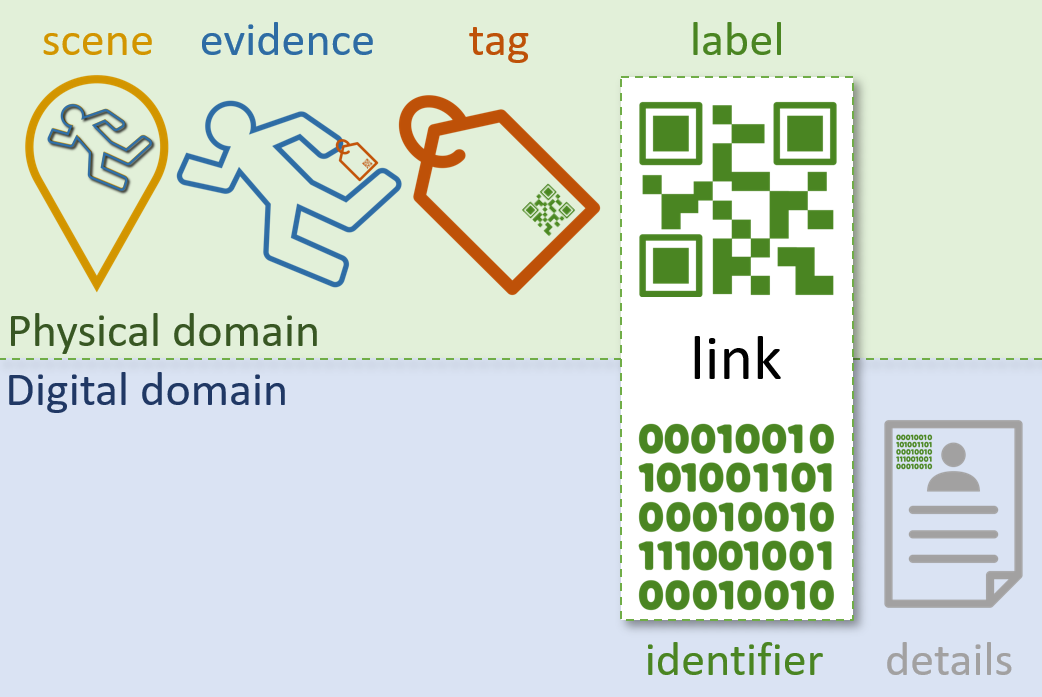}
\end{center}

\subsection*{Key Points:}
\begin{itemize}
    \item Disaster victim identification (DVI) processes require us to link physical evidence and digital information
    \item While technology could improve this linkage, experience shows that technological ``solutions" are not always adopted in practice
    \item Our study of the practices, preferences and contexts of Australian DVI practitioners suggests ten critical considerations for these technologies
    \item We review and evaluate 44 candidate technologies against these considerations and highlight the role of human factors in adoption
\end{itemize}

\newpage

\section{Introduction}

The 2004 Indian Ocean earthquake and tsunami (known colloquially as the Boxing Day Tsunami) is the largest single-incident disaster victim identification (DVI) operation conducted to date.  The scale of the incident and the diverse international response created significant challenges in managing human remains and associated information.  These challenges included allocating unique identifiers to dead bodies and human remains, tracking the location of physical remains as they progressed through DVI processes, and associating and communicating new information about the remains. This leads to the provocation for this paper: \emph{by linking physical evidence to digital information, how can technology help us tag and track the victims of disaster?}

\begin{figure}
    \centering
    \includegraphics[width=\textwidth]{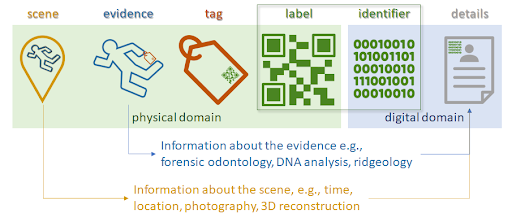}
    \caption{The chain of links between a \emph{scene} containing physical evidence and digital information about that evidence. A \emph{tag} is attached to the \emph{evidence} and a \emph{label} is attached to the tag. This label embodies and encodes a unique digital \emph{identifier} which is then linked to further \emph{details} about the evidence (e.g., forensic odontology, DNA analysis, ridgeology) and scene (e.g., time, location). The label and identifier create the link between the \emph{physical domain} (the real world) and the \emph{digital domain} (the world of information). This paper focuses on technologies related to the linkage of physical and digital information (inside the shaded background), not the technologies used to measure, characterise or analyse the evidence or scene (outside the shaded background).}
    \label{fig:graphical_abstract}
\end{figure}

DVI is information intensive, and the digital domain facilitates information work through lossless storage and communication (including copying) as well as manipulation, analysis, search, and retrieval.  However, that information is about unique physical entities: human beings, human remains, disaster scenes, and other physical evidence. To take advantage of the digital domain, we need to create reliable links from these unique physical entities to information about them (a process we refer to as ``tagging'', see Figure~\ref{fig:graphical_abstract}) along with associated data entry processes. 

Various information standards exist to organise how this linkage is expressed. For example, the Global Standard 1 identification system (GS1) \cite{gs1_gs1_nodate} provides a range of open and international standards for tags and labels (or ``physical data carriers'')\footnote{ 
    GS1 standards underpin identification of products, services and assets worldwide in a range of industry sectors, most prominently retail supply chains (e.g., as the familiar product barcode) but also in healthcare and humanitarian logistics.}. 
GS1 standards enable automated creation of unique identifiers which can then be used in a broad range of human and machine readable label formats. Labels can carry other information in addition to the unique identifier, as well as a mix of publicly and privately readable information.

\begin{figure}
    \centering
    \includegraphics[width=\textwidth, trim={2cm 5cm 2cm 4cm}, clip]{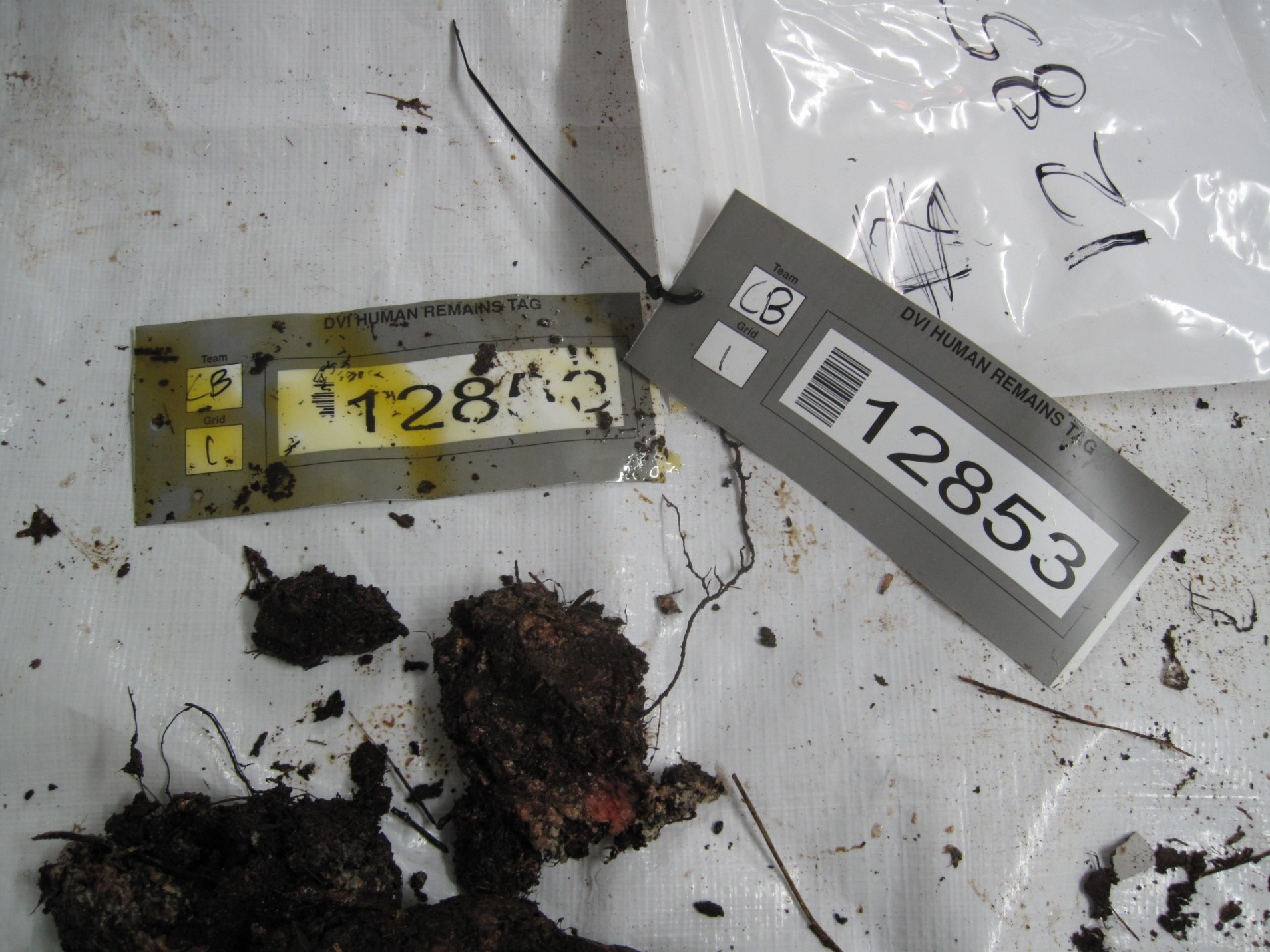}
    \caption{The link between physical evidence and digital information can be fragile: aviation fuel has damaged the printing (i.e., the label) on the left tag. (Source: Peter Ellis.)}
    \label{fig:fuel_label}
\end{figure}

Yet links between physical and digital domains can be \textit{fragile}: Figure~\ref{fig:fuel_label} shows a situation where a tag’s printed label has been partly dissolved. If a label is destroyed, so too is the ability to connect physical remains to information about them. With a small number of fatalities, this link might be reestablished if noticed in time, but in a large multi-fatality event, a corrupted label may mean that a victim’s identity can never be confirmed.  Even with intact tags and labels, the link between physical evidence and the attached information about it might still be overlooked, as happened in the failed repatriation of Australian soldier Private Jacob Kovco in 2006 when ``\dots no one examined the body with sufficient care to notice a cardboard tag attached to the hand bearing the name Juso Sinanovic'' \cite{cosson_inquiry_2006}.

In this study we seek to assist DVI practitioners and others whose work relies on physical evidence linked with digital information.  As Elmasllari and Reiners showed with triage \cite{elmasllari_towards_2017}, technological ``solutions'' are not always welcome in practice. So, our study seeks to help avoid promoting technology that is unlikely to be adopted; hence our focus is on understanding existing work practices and contexts.  To ground our suggestions in the lived experiences of DVI practitioners we ask, not only about what technologies can help, but also \textit{what systematic factors impact the tagging and tracking process?} To answer these questions we drew upon the experience of Australian DVI practitioners.

\subsection{Our contribution}
Rather than creating a system into which work practices must fit, we focus on how technology might lubricate the many joints of established DVI practices.  We conducted interviews and workshops with Australian DVI practitioners to understand the context in which tagging happens. From this data we identified four themes which affect the use and uptake of technology in the DVI context:

\begin{itemize}
    \item protocols and training
    \item stress and stressors
    \item the plurality of information capture and management systems
    \item practicalities and constraints.
\end{itemize}

Using these themes, we then identified key attributes of technologies that affect their use and usefulness in a DVI context.  Finally, we considered candidate technologies that may improve the efficiency and quality of DVI work practices, identifying their key attributes with careful consideration of their costs, benefits and contexts of use. We intend for this overview of candidate technologies to be useful to decision makers and practitioners of DVI operations who seek to improve the efficiency of their workflow, increase the auditability and security of information, and build greater resilience into their systems. Some of the technologies are readily available and customisable, others are emerging and worth closer attention in the near future. 

We have approached our study with Australian first responders and DVI practitioners in mind. These professionals face large-scale disasters relatively infrequently; daily, they deal with much smaller numbers of deaths in which victims need to be identified. This day-to-day work is where DVI protocols, procedures and practices become embedded in operational culture, an important point to remember in designing new technology: if it is readily adopted for use on a daily basis, it is more likely to be employed in extraordinary circumstances. Mercifully infrequent, yet highly diverse, disasters are always extraordinary \cite{taylor_taxonomy_1987}.

\subsection{DVI practice}
Australia follows globally accepted DVI processes, developed by the International Criminal Police Organization \cite{interpol_dvi_working_group_interpol_2018}, currently with 194 member countries. These protocols enable positive, credible identification, using data collected from the scene, and from antemortem and postmortem investigations \cite{interpol_disaster_nodate}. INTERPOL provides a paper-based system to support these protocols, with antemortem, postmortem and identification forms printed on different colours to avoid misinterpretation \cite{sweet_interpol_2010}; KMD provide a commercial system fully integrated with INTERPOL’s DVI forms \cite{noauthor_kmd_nodate}. INTERPOL’s forms and the KMD DVI System International focus primarily on managing and reconciling information about human remains. 

With care to ensure alignment with INTERPOL’s protocols, the International Committee of the Red Cross, World Health Organisation, Pan American Health Organisation and International Federation of the Red Cross have developed and revised the Management of Dead Bodies After Disasters Mmanual \cite{cordner_management_2016}. Designed for situations where forensic services are scarce, this manual has also been well received in countries with well-resourced, highly advanced forensic services, including Australia. 

INTERPOL describes the identification processes of DVI work in four phases (there is a fifth ``debrief" phase focused on quality control and training):
\begin{enumerate}
    \item \textbf{Data collection} at the scene;
    \item \textbf{Ante-mortem}, in which all background information that may aid identification is collected;
    \item \textbf{Post-mortem}, in which information is gathered from the remains;
    \item \textbf{Reconciliation} of scene, ante- and post-mortem information into a positive identification.
\end{enumerate}
Technology already plays an important role in each of these phases; research into new technologies and new applications of technology is ongoing.

\subsection{Examples of technology in DVI practice}
Current research into technology for DVI focuses on data collection at the scene, e.g., digital capture of fingerprints \cite{johnson_digital_2018}; pathology technology to assist the post-mortem phase \cite{sidler_use_2007}; or data management and matching at the reconciliation phase  \cite{hofmeister_icrc_2017}. Fewer studies focus on the \textit{linkage} of physical remains and digital data, or related problems such as the tracking of living victims of disaster. Pate’s \cite{pate_identifying_2008} review identified technologies relevant to tracking both the living and the dead victims of disasters. Those relevant to DVI included barcodes; biometrics (e.g., fingerprint and retinal scanners); forensic odontology;  mobile communications; radio frequency identification (RFID); smart cards; and the Internet. 

RFID has been used by morgue workers to quickly and accurately identify remains during Hurricane Katrina \cite{associated_press_chips_2005,gadh_radio_2006}. Passive RFID chips were implanted in shoulders or placed inside remains bags, and used in combination with an adhesive barcode showing the same RFID number.  A study of the effectiveness of RFID implantation of disaster victims was carried out on the victims of the 2004 Indian Ocean Tsunami, in response to the loss of information on paper tags due to exposure to formaldehyde \cite{meyer_implantation_2006-1}. Challenges included where to implant an RFID to ensure it stayed with the remains over time (they decided on the nasal cavity), and the power needed for scanners to read passive RFID through body bags. The authors felt that conventional tagging processes were adequate with manageable numbers of bodies in average condition, but with very large numbers of bodies in advanced stages of putrefaction, RFID tagging offered significant advantages. Unlike the Hurricane Katrina operation, no mention was made of RFID chips placed inside bags containing incomplete remains, suggesting that this approach is most appropriate when bodies remain intact, otherwise they would need to be supplemented with additional labelling. 

Most explorations into new technology to enhance tagging and tracking have been for medical casualty triage, where the aim is to preserve life. Notable among these is the Wireless Internet Information System for Medical Response in Disasters (WIISARD) to better manage tracking in mass casualty events \cite{lenert_design_2011}. WIISARD was developed using rapid prototyping---leveraging and building on existing relationships with first responder units---and implemented via training exercises such as simulated mass casualty events. This approach was favoured because ``people don’t know what they want before they’ve experienced it'' \cite{lenert_notitle_2021}. This modular system used a scalable network; a database; water-resistant intelligent triage tags providing communication, an electronic medical record and RFID; a sensor platform to monitor the vital signs of casualties; a wireless handheld device allowing responders to record information; and management systems encompassing supervisors of groups of responders and a command centre. In test conditions, WIISARD greatly reduced missing and duplicated patient identifiers \cite{lenert_design_2011}. The authors acknowledge the complexity of the system to build \cite{lenert_intelligent_2005}, and no recent evidence can be found as to its adoption. This suggests to us that factors other than users’ technical proficiency and the potential of the new technology need to be considered in the adoption of new technologies into existing work practices. 

\subsection{Adoption (or rejection) of new technology}
Understanding why new technologies are adopted or rejected has shifted from an emphasis on attitudes and perceptions of usability and usefulness towards interrogating the social practices around technology use \cite{orlikowski_using_2000}. The latter approach recognises that technology is not just used within existing social structures, but that its use can change these structures: organisations can use technology to reinforce and preserve the status quo, or transform it.  Orlikowski’s modelling of the structural consequences of different types of institutional and technological conditions suggests that collaborative or cooperative endeavours will readily adopt tools to increase effectiveness and cooperation. 

The experience of technology use in disaster victim triage provides important lessons for the design and adoption of technology to support DVI. Elmasllari and Reiners \cite{elmasllari_towards_2017} identified 17 different electronic triage systems, pointing out that ``none of these systems [was] in use by first responders, who instead prefer paper-based tools from the 1960s''. This is a salutary warning to consider carefully the different needs, preferences and situations of those who would have to adopt (or adapt to) new technology. The challenge is not to develop digital technology that could \textit{potentially} help us tag and track the victims of disaster, but to understand first the characteristics of technology that would \textit{actually} be useful, usable, and used in DVI practice.

Elmasllari and Reiners \cite{elmasllari_towards_2017} identify several factors impacting the adoption of e-triage systems like WIISARD \cite{lenert_design_2011}.  They provide practical suggestions such as allowing freedom to move between manual and automated systems, and freedom to enter data in different orders to accommodate dynamic conditions and the \textit{ad hoc} nature of response. This flexibility contrasts with the use of more rigid DVI protocols, but acknowledges the need ``not only for graceful degradation but also useful degradation to a well-known, minimum-functioning state''  in both medical and DVI responses (\cite{elmasllari_towards_2017}, p. 1588). Other recommendations include the need to avoid ``information flood''; designing for the physical environment and restrictions of the different roles of responders; and designing interactions and interfaces for extreme stress. 

In another domain-adjacent example, Yao, et al.’s \cite{yao_adoption_2012} review of the use of RFID in healthcare identified issues impacting adoption, including costs, technological limitations, privacy concerns, interference with other equipment, and a lack of global standards. The authors make several suggestions to increase the likelihood of adoption, including that a return on investment analysis be carried out, and that interoperability be improved through pairing RFID tags with barcodes and existing wireless networks.

What is the relative importance of these different barriers to adoption? Reddick’s \cite{reddick_information_2011} survey of US state government emergency departments found that financial constraints and a lack of support from elected officials ranked highest in terms of barriers to the adoption of IT in emergency management---more so than issues with interoperability or lack of expertise. However, within DVI contexts, an investment analysis of new technologies should also include the costs (e.g., reputational) of not addressing the challenges of tagging and tracking, particularly in international contexts. The use of multiple languages on forms, partial or copied forms, relabelled remains, or the use of multiple identifiers for the same body have all been reported as problems in real disasters \cite{byard_potential_2010})---each greatly increases the chances of releasing the wrong bodies to grieving families.

\section{Method}

This study sought to learn how the lived experiences of DVI practitioners affect technology requirements and adoption, and to present a set of candidate technologies for consideration. The investigation occurred in two parts:
\begin{enumerate}
    \item A human-computer interaction (HCI) led investigation into DVI practices in the Australian context to provide an understanding of the needs and constraints of the practitioners. This involved interviews and a workshop.
    \item An series of discussions amongst the authors reflecting upon the insights of the first part of the investigation, as well as extant literature, to determine what technologies and considerations to present. The authors have a combined expertise across electronics, data science, cybersecurity, human-computer interaction, and forensic pathology. 
\end{enumerate}
Part 2 was informed by thematic analysis of the data collected in Part 1. Thematic analysis was conducted following the process outlined in Braun and Clarke \cite{braun_using_2006}. This involved close reading of interview and workshop transcripts, followed by iterative development of codes into higher level themes. These themes were initially independently developed by Author 1 and Author 2. This was followed by discussion with Author 6 and insights were integrated into the iterative coding process until agreement was reached. Thematic analysis provides a ``rich and detailed, yet complex account of data'' (\cite{braun_using_2006}, p. 78), without discarding generalisability and, as such, is appropriate when seeking to learn how work practices might influence which types of technological supports are most likely to be adopted. 

\subsection{Scope}
We consider civilian contexts only (not military settings) and we concentrate on situations where there is likely to be uncertainty about the identity of the victim (e.g., where death has occurred outside a hospital environment). We focus on the linkage between physical and digital domains, not the actual measurement, characterisation or analysis of evidence.

\subsection{Participants}
In Australian DVI operations, scene retrieval of remains and DVI coordination may be conducted by various state policing services or coordinated by Australian Federal Police, while pathologists and mortuary managers are drawn from various medical and academic institutions. DVI procedures are most often enacted in events with low casualty numbers. This serves as ongoing training for participation in larger operations, which require greater levels of coordination and resourcing. Within this study we spoke to participants whose experience of DVI varied across numbers of casualties, environments, and disaster types. However, the DVI processes and procedures remained constant, albeit with greater challenges in larger-scale disasters. 

Ethical approval was gained from the Queensland University of Technology Ethics Committee (Approval \#1900000363) to conduct low-risk human research. Individual interviews and a workshop were conducted to learn about the DVI contexts, practices, challenges, and actors. Interviews were conducted with seven DVI practitioners, and the workshop with three of the same participants (all members of the Queensland Police Service (QPS)). (See Table~\ref{tab:participants} for participant information.)

Interview questions focused on the participants' backgrounds and practices, with specific questions regarding DVI data management, technologies used, the agencies they interacted with, and the physical and cultural environments they have conducted DVI in. Interviews ranged from 32 to 68 minutes in duration (M = 42min, SD = 14min).

\begin{table}[t]
    \centering
    \begin{tabularx}{\textwidth}{c p{0.9\textwidth}}
        \toprule
        ID & Background \\
        \midrule
        P1 & Mortuary Manager at a major Australian medical institute. \\
        P2 & Forensic odontologist. \\
        P3 & Australian Federal Police operative. \\
        P4 & Senior Sergeant (QPS) \\
        P5 & Chair ADVIC (ANZPAA (Australia New Zealand Policing Advisory Agency) DVI Committee) and Senior Sergeant in Australian Federal Police. \\
        P6 & Sergeant (QPS) \\
        P7 & Senior Sergeant and DVI Coordinator (QPS) \\
        \bottomrule \\
    \end{tabularx}
    \caption{Details of the participants in this study.}
    \label{tab:participants}
\end{table}

Additionally, a workshop was held with three senior members of the QPS (P4, P6, P7) to better understand DVI processes at the scene. This was conducted at QUT, over 1 hour 38 minutes, recapitulating what the researchers had learned from the interviews to give the workshop participants a chance to correct and add detail. This was followed by the presentation of a technology probe: a mock-up of a physical tag and associated digital system, which sparked discussion about the applicability of its components. The participants then conducted a recreation of scene procedures, with each participant playing one of the following roles: scene coordinator/recorder, body handler, and photographer. Quotes from the workshop are indicated by ``(WS)'' as it was not possible to identify individual responses. 

All audio material was transcribed. This, plus email correspondence, formed the dataset.

\section{Findings}
We have organised our findings in three parts. First we present what we learned from participants about systemic factors affecting DVI practices in Australia: this helps us identify issues that are important in developing and deploying new technology. Second, we use these insights and our combined domain expertise (in electronics, data science, cybersecurity, human-computer interaction, and forensic pathology) to set out a range of important considerations for technologies that seek to enable better linkage of physical remains and digital data. Third, we present a range of plausible candidate technologies and review how their attributes address these considerations.

\subsection{Systemic factors affecting DVI practices in Australia}

Qualitative analysis of the dataset allowed us to identify systemic factors that affect the quality of DVI practices. We identified four main themes in our participants’ responses.

\subsubsection{Protocols and training}
Interviewees alluded to a strong faith in existing processes when describing INTERPOL processes. As P6 stated: ``you’re supposed to align\dots your procedures to the INTERPOL\dots it's not like we can just decide we're going to do it this [other] way. You shouldn't be doing that.'' This faith was based upon a success rate that was ``scientifically proven'' (P7), however, it could also produce an attachment to procedure that could lead to ``checking the fingerprints even though we know they haven't got hands'' (P5).

Practices are scaled up from the local to larger DVI operations through ongoing local-level training and adherence to INTERPOL processes. This was apparent when the idea of having one tagging system for the local level and one for the international level was broached: 

\begin{quote}
It goes against the process. I can understand the possible need for it, but the protocols would have to be written so everybody understands it. You don't want people coming in to go, ``Oh, you've learned this way forever and a day, but now we're doing this”---P7
\end{quote}

Similarly, changes were initiated at the state level before advocating national change: 

\begin{quote}
\dots[the Scene Recovery Booklet] is a Queensland document\dots we've recently modified it\dots last week, went to the annual Australian conference for everybody to look at\dots because they want to try and use it everywhere across Australia and New Zealand.---P6
\end{quote}

Potential changes to protocols were identified during training exercises, as well as regular operations. For example, QPS’s Scene Recovery Booklet was examined in light of its use:

\begin{quote}
\dots for the last two or three years, we've run scene phase training, and from there, the coordinators sit down and actually look at it\dots And basically every year we review. We'll sit down and just go, is it working? Is it not working? And we get feedback from the whole squad on that. So they're the ones that actually go and do it. We're testing them, but then it's like, ``Oh, what do you think of the book? Is there anything that needs changing?"---P6
\end{quote} 

However, within Australia, there are considerable differences in the way that DVI staffing is arranged across the different states, with subsequent impacts on training. For example, P6 reports:

\begin{quote}
\dots so Victoria for instance\dots their antemortem and their recon, is done by the detectives. New South Wales, if you're a scene-of-crime officer, you're automatically a DVI officer\dots  Up here [Queensland], we're totally voluntary. We ran a course this year. I think we're seeing fifty-five. So we may still have the lowest numbers. But then we've got fifty-five people training at every phase, where other states can't say that. I think WA [Western Australia] went away from that, and they've come back to it. So they're starting to train their staff up. Tasmania, their scene team, and this was two or three years ago when I did a coordinators’ course, their scene team hadn't been together for five years\dots And our guys are trained every year in every phase.---P6
\end{quote}

The implication for DVI work practices is that routine, local operations can exert powerful forces; even though change may be extremely globally beneficial in unusual situations, the benefits to local, routine operations will strongly influence new technology adoption. Importantly, proposed technologies should facilitate existing processes and procedures without significant disruption, and in ways that still allow for human oversight. These could be best introduced and tested during training exercises. 

\subsubsection{Stress and stressors}
Participants reported engaging in repetitive activities that were physically and mentally draining. These could occur across the DVI phases but were particularly noticeable in the postmortem phase. For example, P2 recounts the intensity of the smell within the morgue during a mass disaster, which took time to acclimate to:
\begin{quote}
\dots if you leave, you've got the same experience going back in, the next time. So, the urge not to leave is really, really strong\dots if you're working for six or eight hours at a go, you're full of adrenaline, you're pumping, you have no sense that anything is a mess\dots up becomes down, back becomes front, left becomes right, and you don't notice. And the errors compound the longer you go.
\end{quote}

Similarly, large-scale disasters present challenges to mortuary staff who must ``\dots spend hours searching for bodies in fridges with racks, and racks, and racks of them'' (P1). Also reported by P1, a solution of a whiteboard with a schematic of the freezer unit and magnets that corresponded to remains’ IDs was used to great success, and could be potentially be translated to a more automated system. Beyond tracking human remains, there is also a need to identify what tests have been or need to be carried out---a problem magnified by large-scale operations.

This study was unable to gain access to the personnel tasked with data transcription, i.e., digital data entry of paper scene recovery booklets. However, data was reported as being manually transcribed into data matching software (Plass Data):

\begin{quote} 
\dots we're happy to wear that basically per file: an hour data entry for an AM, an hour data entry for a PM, because the ability to do the advanced searching and the functionality of it to reduce our end paperwork without having to retype in numbers---the benefits outweighed that time.---P4
\end{quote}

Direct importation of this data was hindered by reliance on paper-based scene data collection (for reasons explored below). Regardless, the mental load of DVI response within a complex and changing environment---or even a simpler one---produces the need for easily understood processes and technologies; as P1 relates: ``\dots in a disaster we keep it simple… we don’t want a system that we’re using for the first time.''

Training offers one solution for improved quality assurance by reducing both mental load and reinforcing clear protocols.  Further to this, technologies that can reduce repetition, reduce mental load (or at least not add to it), and improve the ability to track or locate remains in mortuaries are likely to be appreciated. 

\subsubsection{Plurality of information capture and management systems}
Across Australia, DVI involves a range of information systems.  In Queensland, for example, data is always entered into the state’s Forensic Register and selectively shared into the National Missing Person and Victim Service (NMPVS). Consequently, there is no guarantee that state-level information will be shared nationally. Since missing persons can be considered ``a mass disaster happening over a long period of time'' (P2), and since information about missing persons is used in antemortem DVI investigations, ways to facilitate greater information exchange between databases would be valuable. More aligned systems, particularly ones in which data can be immediately entered into a shared system, could also produce greater efficiencies both within and across DVI phases.

P5 imagines an idealised system in which:
\begin{quote}
\dots  we could input data from the scene directly into a database that would capture it and provide information through to the next, the other phases, which will inform him of all the circumstances and\dots opportunities. For example, if we are getting a lot of information or we were at a scene that had a particular type of remains that lends itself to a particular primary identifier, then that's something we can prepare for both in the PM phase and the antemortem phase.
\end{quote}

However, identifiers may change or multiply throughout the DVI process. P1, a pathologist, notes that errors at the scene may need to be corrected at the PM phase:
\begin{quote}
Ideally you tag once at the scene and that stays as a tag and that doesn't change. It has consistency. But there are occasions in which you can have a body bag, and a scene person would've put three people in one body bag. So that means that\dots if you've got three knees in there, one knee has to come out or the two knees have to come out, so then we create new case numbers in the mortuary.
\end{quote}

Multiple databases increased confidence that data was backed up and could be accessed if one system failed. Similarly, individual paper-based data capture (i.e., one scene recovery booklet for each remain) was seen as a safer way to capture information than a system that relied on a single device whose failure could be catastrophic. As P7 relates: ``I would much rather have handwritten notes. Then we can scan it\dots If something were to happen to that device on the way, you've lost all your data.'' A similar concern was voiced regarding the idea of all the scene information being held on a RFID tag:
\begin{quote}
It's always got to be linked to some other central computer system for us as well. Can't be just attached to a body. I always find body tags\dots go missing. As long as it's all connected to some sort of central mainframe and central location.---P1
\end{quote}

The need for data entry and database design that reduces the frictions of data collection and improves information exchange between databases is indicated. However, the application or design of new technologies will also need to be respectful of separate jurisdictions and their need to maintain data security.  Finding opportunities to introduce data backup or partitioning of data will also sustain data continuity and confidence in the use of new technologies.

\subsubsection{Practicalities and constraints}
DVI procedures at the scene and post-mortem phases entail interactions with diverse physical environments and remains. Reasons for avoiding new technologies reflected concerns that they would add to the effort: 
\begin{quote}
 \dots we had a 1km hike to the crash site and we went down 900 meters\dots if I've got to lug something with me apart from my equipment to bring the body out, you're adding that on. Like, I don't know what you'd need to take, but how do we test that whatever we tagged them with is actually working? Do we need a portable something or other? And then how much does that cost? And then where’ve we got to put it, or where else do we need it?---P6
\end{quote}

Typically, the only electronic devices used at the scene would be cameras, used by the photographer who ``pulls on rubber gloves---but they're clean. The photographer doesn't touch the body'' (WS). Relatedly, personal protective equipment hampered the use of touch-screens within post-mortem data collection, and the presence of bodily fluids hampered all direct data collection, requiring information to be communicated verbally to an assistant to record it (as reported by P2). This was another reason why paper scene recovery booklets were preferred:

\begin{quote}
All that recording on the book, the recording booklet is actually done in the scene on hardcopy papers. We've looked at medical Toughbooks, that sort of stuff, and we know that they're fine for operating theatres. They're not necessarily good for being dropped on the side of the hill. Ruggedized Toughbooks won’t like being wiped over with bleach and chemicals and everything. You need to get all human remains out of them.---WS
\end{quote}

Identification tags had to be robust to liquid, chemical and mechanical degradation, yet able to readily capture and communicate information. At a minimum, they needed to convey the site from which the remains were retrieved, the team conducting the retrieval, and a unique identifier which could be visually read by a human and optionally read by a machine. For this, water-repellent paper tags that could be easily printed were preferred and used, with a barcode and nine digits for identification purposes. 

While options such as video capture, dictation, and speech-to-text technology were mentioned as possible ‘hands-free’ solutions to data entry, the idea of capturing interpersonal communications, particularly at the scene, evoked mixed reactions: 
\begin{quote}
Due to the nature of a lot of conversations that go on, we don't want to be video recorded\dots If it was a live, just a direct feed with no recording, we could possibly look at that ---WS
\end{quote}

Speech-to-text and other methods (e.g., live streaming with no saved recordings) could be deployed in ways that avoid this concern. 

DVI may also be competing for resources with other priorities in larger organisations. In the initial disaster response, external pressures (e.g., media, political) can generate greater resourcing (e.g., staffing), but as the DVI process stretches out and other priorities emerge, the reconciliation phase may be under-resourced. Practitioners may also face dual workloads: ``my business as usual work is backlogging while I'm off doing a DVI job'' (P4). 

While some technological solutions may offer potential improvements to efficiency by reducing repetitive data entry, we need to consider how practical concerns of robustness, accessibility and affordability overlap with social concerns (i.e., interpersonal trust and communication), affecting technology use and adoption. Stakeholders also have to be convinced that the benefits of change offset other costs, including the opportunity cost of investing in that technological change rather than another priority.

\subsection{Key considerations and technology attributes}

Synthesising the insights gained from the interview and workshop investigation of DVI practices with the combined expertise of the authors, we developed a list of ten high-level considerations to keep in mind while assessing the potential of technologies for local or global adoption in DVI (Table~\ref{tab:considerations}).

These general considerations can be linked to specific technology attributes or features that are meaningful to DVI practice. Table~\ref{tab:techattributes} groups these attributes and relates them to the considerations listed in Table~\ref{tab:considerations}. Together, these Tables help organise our review of candidate technologies in a way that relates to what we learned from our study participants. 

\begin{table}[tbp]
    \centering
    \begin{tabular}{cl}
        \toprule
         & Consideration \\
        \midrule
        1 & Fit with existing processes, procedures and practices \\
        2 & Efficiencies in data collection and sharing \\
        3 & Enhancement of inter-task cooperation \\
        4 & Digital information sharing \\
        5 & Ease of tracking and locating remains \\
        6 & Quality assurance and oversight\dots trust \\
        7 & Acceptability to separate jurisdictions \\
        8 & Accessibility and affordability \\
        9 & Robustness to environmental conditions and uses \\
        10 & Security and auditability  \\
        \bottomrule \\ 
    \end{tabular}
    \caption{These ten points emerged from discussions with study participants as important high-level considerations in assessing technologies to better link physical evidence and digital information for DVI.}
    \label{tab:considerations}
\end{table}

\begin{table}[p]
    \centering
    \small
    \rowcolors{2}{white}{gray!20}
    \begin{tabularx}{\textwidth}{nw}
\textbf{Is/does the technology...?}                      & \textbf{In other words...}                                                \\
\hline
Readily \textbf{human readable}?                         & Can a human easily read text on the tag, device or system?                \\
Readily \textbf{human writable}?                         & Can a human easily write text on the tag, device or system?               \\
Readily \textbf{machine readable}?                       & Can computers read information from the tag, device or system?            \\
Readily \textbf{machine writable}?                       & Can computers write information to the tag, device or system?             \\
Involve a \textbf{human in the loop}?                    & Are humans able to check and intervene in the system?                     \\
\textbf{Automatable}?                                    & Can the process be fully automated?                                       \\
\textbf{Conveniently usable} for humans?                 & Is the system especially convenient for humans to use?                    \\
\textbf{Internationally interpretable}?                  & Can information be understood no matter what languages you speak?         \\
\textbf{Degrade gracefully}?                             & If the system is damaged, is it still useful to some extent?              \\
\textbf{Robust} to environment and failures?             & Can the system withstand harsh operating conditions?                      \\
\textbf{Usable} in many contexts?                        & Is the system useful in many different situations?                        \\
Use \textbf{readily available} equipment/consumables?    & Are the necessary equipment or consumables easy to obtain?                \\
\textbf{Compliant} with widely adopted standards?        & Is the tag, device or system standards compliant?                         \\
\textbf{Interoperable} across jurisdictions?             & Can the system be used across different state or federal jurisdictions?   \\
\textbf{Inexpensive}?                                    & Is the tag, device or system relatively cheap to buy and use?             \\
\textbf{Enable items to be located} in physical space?   & Does the system help us find specific items?                              \\
\textbf{Support additional features} or special uses?    & Can the functionality of the tag, device, system be extended?             \\
\textbf{Provide high information transfer} rates?        & Does the system allow rapid data transfer?                                \\
Support \textbf{decentralised} control of functionality? & Does the system avoid the need for control by a central authority?        \\
\textbf{Provide redundancy}?                             & Can a single task be accomplished by more than one device?                \\
\textbf{Highly available}?                               & Can services be accessed when and where they are needed?                  \\
Preserve \textbf{confidentiality}?                       & Are unauthorised  parties prevented from reading data?                    \\
Ensure \textbf{data integrity}?                          & Can the system ensure data has not been altered since it was entered?     \\
Enable transaction \textbf{audit}?                       & Do transactions create records that can later be checked?                 \\
Enforce \textbf{authentication}?                         & Are users required to identify themselves to gain access to the system?   \\
Enforce \textbf{authorisation}?                          & Can permissions to perform transactions can be granted or revoked?        \\
Enforce \textbf{access control}?                         & Can authorisation be granted to read or write data?                       \\
    \end{tabularx}
\vspace{2ex}
    \caption{Technology attributes that relate to important considerations in DVI practice (see Table~\ref{tab:considerations}), e.g., is the technology human readable? Does the technology enforce access control?}
    \label{tab:attributes}
\end{table}

\setlength{\tabcolsep}{2pt}
\begin{table}[p]
    \centering
    \small
    \rowcolors{2}{gray!20}{white}
\begin{tabular}{l!{\color{white}\vrule}c!{\color{white}\vrule}c!{\color{white}\vrule}c!{\color{white}\vrule}c!{\color{white}\vrule}c!{\color{white}\vrule}c!{\color{white}\vrule}c!{\color{white}\vrule}c!{\color{white}\vrule}c!{\color{white}\vrule}c}
\textbf{Is/does the technology…? }                         &
\rot{Fit with existing processes, procedures and practice} &
\rot{Efficiencies in data collection and sharing         } &
\rot{Enhancement of inter-task cooperation               } &
\rot{Digital information sharing                         } &
\rot{Ease of tracking and locating remains               } &
\rot{Quality assurance and oversight… trust              } &
\rot{Acceptability to separate jurisdictions             } &
\rot{Accessibility and affordability                     } &
\rot{Robustness to environmental conditions and uses     } &
\rot{Security and auditability                           } \\
\hline
Readily \textbf{human readable}?                            & \T &    &    &    &    & \T &    & \T &    &    \\
Readily \textbf{human writable}?                            & \T &    &    &    &    & \T &    & \T &    &    \\
Readily \textbf{machine readable}?                          &    & \T & \T &    &    & \T &    &    &    &    \\
Readily \textbf{machine writable}?                          &    & \T & \T &    &    & \T &    &    &    &    \\
Involve a \textbf{human in the loop}?                       & \T &    &    &    &    & \T &    &    & \T &    \\
\textbf{Automatable}?                                       &    & \T &    &    &    & \T &    &    &    &    \\
\textbf{Conveniently usable} for humans?                    &    & \T &    &    &    &    &    &    &    &    \\
\textbf{Internationally recognizable and interpretable}?    &    &    &    &    &    &    & \T &    &    &    \\
\textbf{Degrade gracefully}?                                &    &    &    &    &    & \T &    & \T & \T &    \\
\textbf{Robust} to environment and failures?                &    &    &    &    &    &    &    &    & \T &    \\
\textbf{Usable} in many contexts?                           &    &    &    &    &    &    &    & \T & \T &    \\
\textbf{Use readily available equipment} or consumables?    &    &    &    &    &    &    &    & \T &    &    \\
\textbf{Compliant} with widely adopted standards?           &    &    &    &    &    &    & \T &    &    &    \\
\textbf{Interoperable} across jurisdictions?                &    &    &    &    &    &    & \T &    &    &    \\
\textbf{Inexpensive}?                                       &    &    &    &    &    &    &    & \T &    &    \\
\textbf{Enable items to be located} in physical space?      &    &    &    &    & \T &    &    &    &    &    \\
\textbf{Support additional features} or special uses?       &    &    &    &    & \T &    &    &    &    &    \\
\textbf{Provide high information transfer} rates?           &    & \T & \T & \T &    & \T &    &    &    &    \\
Support \textbf{decentralised} control of functionality?    &    &    &    & \T &    &    & \T &    & \T & \T \\
\textbf{Provide redundancy}?                                &    &    &    &    &    &    &    &    & \T &    \\
Services are \textbf{highly available}?                     &    &    &    &    &    &    &    &    & \T &    \\
Preserve \textbf{confidentiality}?                          &    &    &    &    &    &    &    &    &    & \T \\
Ensure \textbf{data integrity}?                             &    &    &    &    &    & \T &    &    &    & \T \\
Enable transaction \textbf{audit}?                          &    &    &    &    &    & \T &    &    &    & \T \\
Enforce \textbf{authentication}?                            &    &    &    &    &    &    &    &    &    & \T \\
Enforce \textbf{authorisation}?                             &    &    &    &    &    &    &    &    &    & \T \\
Enforce \textbf{access control}?                            &    &    &    &    &    &    &    &    &    & \T \\
\end{tabular}
\vspace{2ex}
    \caption{The relationship between different technology attributes (rows) and considerations (columns) identified as important by study participants (see  Tables~\ref{tab:considerations} and~\ref{tab:attributes}).}
    \label{tab:techattributes}
\end{table}

\subsection{Candidate technologies to better link physical evidence and digital information}
What follows captures many of the ideas that arose in the course of our study, but is not an exhaustive list. We hope that the range of candidates we describe, and the way we organise our thinking about them will be a useful basis for DVI practitioners and policy makers to consider the merits of different alternatives.

We organise candidate technologies into those that primarily relate to the physical domain, to the digital domain, or whose role is to span both domains (Table~\ref{tab:technologies}).
\begin{itemize}
    \item The physical domain refers to the real world in which physical evidence exists.
    \item The digital domain is where information is represented digitally.
    \item Some technologies straddle these two domains, connecting the physical to the digital, linking or supporting transactions between them. 
\end{itemize}

In comparison to Pate’s (2008) list of technologies for identifying and tracking disaster victims, our categorisation focuses on the linkage of evidence and information and does not consider technologies that assist search and rescue (e.g., robotics) or measuring the evidence (e.g., biometrics, odontology).

Each candidate technology has, or lacks, certain attributes that are relevant to DVI practice. Table~\ref{tab:tech.attr} gives a static view of this information; a searchable Excel spreadsheet version of this, with explanatory comments, is provided as Supplementary Information. 
Some technology-attribute combinations are not especially meaningful: these are left blank. Ticks/crosses indicate where a candidate technology has/lacks a specific attribute. Hollow circles indicate a situation where an attribute may depend on how the technology is implemented, or where further consideration must be given---details of which are given in the searchable Excel spreadsheet.

In the following three sections, we pick out some details to give the reader a useful overview of the technologies and their merits. We suggest that Table~\ref{tab:tech.attr} and the Supplementary Information be used in conjunction with the following overview.

\begin{table}[p]
    \centering
    \small
\setlength{\tabcolsep}{10pt}
\begin{tabular}{lll}
\textbf{Domain}                       & \textbf{Technology Type}                     & \textbf{Candidate Technology}\\                  \hline
\multirow{18}{*}{Physical}            & \multirow{ 6}{*}{Tag technology}             & Flexible tag\\
                                      &                                              & Rigid tag\\
                                      &                                              & Adhesive tag\\
                                      &                                              & Tag with RFID \\
                                      &                                              & Implantable RFID\\
                                      &                                              & Enclosure\\                            \cline{2-3}
                                      & \multirow{ 5}{*}{Label technology}           & Human-interpretable information\\
                                      &                                              & Direct part marking\\
                                      &                                              & Bar code\\
                                      &                                              & Passive RFID\\
                                      &                                              & Active RFID\\                          \cline{2-3}
                                      & \multirow{ 7}{*}{Locator technology}         & Automated locator system\\
                                      &                                              & GPS locator system\\
                                      &                                              & IPS locator system\\
                                      &                                              & Choke point locator system\\
                                      &                                              & Human-powered locator system\\
                                      &                                              & Semi-automated locator system\\
                                      &                                              & Beacon location system\\               \hline
\multirow{ 9}{*}{\makecell{Physical\\to\\Digital}} & \multirow{ 2}{*}{Scene data collection}      & Aerial image capture\\
                                      &                                              & 3D capture\\                           \cline{2-3}
                                      & \multirow{ 4}{*}{Data entry}                 & Automated speech recognition\\
                                      &                                              & Smart-phone sensors\\
                                      &                                              & Machine-readable form elements\\
                                      &                                              & Automated forms\\                      \cline{2-3}
                                      & \multirow{ 3}{*}{Workflow technologies}      & Matching labels\\
                                      &                                              & Workflow metadata\\
                                      &                                              & Automated process logic\\              \hline
\multirow{17}{*}{Digital}             & \multirow{ 2}{*}{Information standards}      & Open identifier standards\\
                                      &                                              & Closed identifier standards\\          \cline{2-3}
                                      & \multirow{ 9}{*}{Information infrastructure} & Local applications\\
                                      &                                              & Remote applications\\
                                      &                                              & Distributed databases\\
                                      &                                              & Opportunistic communication\\
                                      &                                              & Mobile \textit{ad hoc} networks\\
                                      &                                              & Mobile data networks\\
                                      &                                              & Self-managed mobile networks\\
                                      &                                              & Self-hosted computing\\
                                      &                                              & Cloud-hosted computing\\               \cline{2-3}
                                      & \multirow{ 6}{*}{Information security}       & Digital signatures\\
                                      &                                              & Digital certificates\\
                                      &                                              & Encryption\\
                                      &                                              & Access control\\
                                      &                                              & Audit trails\\
                                      &                                              & Block chains\\                         \hline
\end{tabular}
\vspace{2ex}
    \caption{Candidate technologies organised into domains and technology types. These technologies are explained in more detail in Sections~\ref{sec:physical}--\ref{sec:digital}.}
    \label{tab:technologies}
\end{table}

\subsubsection{Physical Domain technologies}
\label{sec:physical}
In the physical domain, the chain of links between the evidence at the scene begins with a tag which is attached to evidence and which bears a label embodying an identifier uniquely associated with digital information about the remains (Figure 1). Locating a labelled tag in the physical domain is important, both in terms of tracking the whereabouts of specific remains, and in retrieving specific remains, say from within a mortuary refrigerator.

Technologies in the physical domain of DVI need to be robust enough to handle heat, cold, moisture, solvents, rough handling; intuitive to use correctly under challenging circumstances; and readily accessible to first responders and other practitioners. Cost will be an important issue when many remains need to be tagged, but we can imagine scenarios where more expensive devices could be reused (e.g., audible or visible “beacons” to help retrieve specific remains in a mortuary).

We organise physical domain technologies into three groups relating to the tag, the label and methods of locating a labelled tag.

\textbf{Tag technology} enables the creation and reuse of physical tags, such as:
\begin{itemize}
    \item \textbf{Flexible tags} on various media (paper, laminated paper, flexible plastic). These still need to be attached to or somehow kept with remains (e.g., with string or cable ties).
    \item \textbf{Rigid tags}, e.g., rigid plastic, as used in ID and credit cards.
    \item \textbf{Adhesive tags} can be stuck on evidence bags or other containers  
    \item \textbf{Tags with RFID} (Radio Frequency Identification) can be made from flexible, or rigid media; adhesive versions are available
    \item \textbf{Implantable RFID}: small glass or plastic capsules containing a passive RFID device. Commonly used for pet     identification.
    \item \textbf{Enclosures}: typically small plastic containers, often waterproof.  Commonly used in logistics with active RFID.
\end{itemize}

\textbf{Label technology} refers to the ways that a unique digital identifier can be associated with a physical tag. Labels can be directly readable or writable by humans, or may need a machine (e.g., scanner, printer) to be read or written. Tags generally use more than one kind of label (e.g., human-readable text plus a machine-readable barcode). 
\begin{itemize}
    \item \textbf{Human-interpretable} labelling can be handwritten or machine-printed directly onto a tag. Information can be encoded in many ways (e.g., numbers, text, symbols, colours) and consideration should be given as to who will have to interpret it (e.g., readers from many different language groups; only staff from a specific mortuary or pathology lab). Thought should also be given as to how easily and accurately this information can be communicated by speech (e.g., over the phone or in a noisy environment) or remembered \cite{shay_correct_2012}.
    \item \textbf{Direct part marking} (DPM) is a term used to describe information physically encoded into the tag material (e.g., by indentations, notches, or holes).
    \item \textbf{Bar codes} are visual, machine-readable representations of data. Most familiar are the 1-D linear stripes used to identify products in retail using the UPC (Universal Product Code), and the 2-D matrix QR (Quick Response) code.
    \item \textbf{Passive RFID}:  Small (centimeter-scale) electronic device, powered by radio-frequency electromagnetic radiation from an RFID reader. Available in many formats, including flexible or encapsulated. Often used in ID cards. 
    \item \textbf{Active RFID}: similar to passive RFID but powered by internal battery which affords greater range and some data storage, in a larger, more expensive format.  Often used in logistics.
\end{itemize}

\textbf{Locator technology} enables labelled tags and associated remains to be spatially located. 
\begin{itemize}
    \item \textbf{Automated locator systems}: Use information gleaned from tags and their interaction with readers to automatically updated database with location information
    \item \textbf{Global positioning systems (GPS)} compare signals from multiple satellites to triangulate position. Lack precision indoors.
    \item \textbf{Indoor positioning systems (IPS)} use transmitters mounted indoors to improve tracking precision.
    \item \textbf{Choke point based systems} use sensors installed in doorways, cabinets or other choke points to detect the passage or presence of a labelled tag.
    \item \textbf{Human-powered systems} manually track and record the location of a labelled tag.
    \item \textbf{Semi-automated systems} require humans and machines to read labelled tags e.g., via handheld scanners.
    \item \textbf{Beacon location systems} help retrieve specific labelled tags by allowing users to trigger an audible or visible response from the tag.
\end{itemize}

Taking \emph{direct part marking} as an example of physical domain technology, we see it can meet some high level considerations by being highly robust (depending on the tag material); it is machine readable, and can be read with common equipment, e.g., mobile phone; and can be inexpensive. However, it is also difficult to create by hand; the equipment needed to create them is not widely available; they are only useful for some tag types; access control is not possible; and while humans can read and compare them they are hard to interpret.

\subsubsection{Technologies spanning physical and digital domains}
It takes time and effort to accurately capture and move information about physical evidence into the digital domain: any technologies that save precious time and effort in DVI are appealing. At all phases of DVI, data collection and entry must also comply with international, national and local protocols; technology has the potential to ease the burden of compliance, and streamline workflows within and across DVI phases.

We have organised technologies that span physical and digital domains into three groups relating to \emph{data collection from the scene}, \emph{data entry}, and \emph{workflow}.

\textbf{Scene data collection} deserves special consideration because, unlike human remains and other pieces of evidence, the scene itself cannot typically be removed or preserved for further analysis. Two technologies stand out as ways to preserve information about the scene and evidence \textit{in situ}, and these are especially important in preserving information when the scene has to be disrupted in the search for further evidence (e.g., by excavation):
\begin{itemize}
    \item \textbf{Aerial image capture}, typically by drones, enables investigators to obtain a plan view of the scene and key locations within it (e.g., the position of remains)
    \item \textbf{3-D image capture}, enables the physical configuration of the scene to be reconstructed.
\end{itemize}

Data entry refers to the transcription of scene, ante-mortem, or post-mortem data into digital formats that increase data longevity and utility. Technologies include:
\begin{itemize} 
    \item \textbf{Automated speech recognition} enables live or recorded speech to be transcribed to text and stored, potentially avoiding manual data entry.
    \item \textbf{Smart-phone sensors}:  modern commodity smart-phones provide an array of sensors and input methods to capture image, video, location, orientation, time, audio and text. Smart-phones can also read and interpret RFID tags, barcodes and QR codes. 
    \item \textbf{Machine-readable form elements} (e.g., barcodes, QR codes, optical mark recognition) enable some automation of optical scanning.
    \item \textbf{Automated forms}, such as fillable PDF forms or dedicated applications on smart-phones, tablets or laptops, can reduce typing and repetition by simplifying what the user has to enter or by re-using existing information.
\end{itemize}

\textbf{Workflow technologies} seek to simplify and facilitate the complex, interdependent phases of DVI processes by ensuring information is communicated within and across phases. 
\begin{itemize}
    \item \textbf{Matching labels} are a simple, practical strategy to ensure information about unique identifiers is reliably propagated through DVI processes. The same identifier is used to label many tags, possibly on different tag media. These labelled tags can be made in advance or on demand.
    \item \textbf{Workflow metadata} carries information about the processes that physical evidence has gone through (rather than information about the physical evidence itself). This could include information about who did what with the evidence, when, where, why and how, so that there is an auditable trail of workflow steps. Workflow metadata is particularly important to indicate related samples (e.g., a tissue sample that has been taken from another set of remains).
    \item \textbf{Automated process logic} uses and creates metadata to guide and control DVI process workflows. This logic could be used to determine, allocate and schedule different process tasks based on the state of the physical evidence (e.g., fingerprinting or odontology may not be indicated for partial remains).
\end{itemize}

Considering \emph{automated speech recognition} as a technology that spans physical and digital domains, it can be useful when hands are busy (e.g., a pathologist during autopsy, a first responder carrying equipment) or typing is slow or inconvenient (e.g., when wearing gloves). However, it requires a quiet environment, and is most convenient when microphones are deployed in the work area for true hands-free operation. This suggests it may have greater uses in controlled environments such as morgues, and fewer in the field.

\begin{table}[p]
    \centering
\begingroup
\setlength{\tabcolsep}{1pt}
\scriptsize
\begin{tabular}{l|cccccccc|ccc|cccc|cc|cccc|cccccc|}
\textbf{Candidate Technology} 
& \rot{Readily human readable?}
& \rot{Readily human writable?}
& \rot{Readily machine readable?}
& \rot{Readily machine writable?}
& \rot{Involve a human in the loop?}
& \rot{Automatable?}
& \rot{Conveniently usable for humans?}
& \rot{Internationally recognizable and interpretable?}
& \rot{Degrade gracefully?}
& \rot{Robust to environment and failures?}
& \rot{Usable in many contexts?}
& \rot{Use readily available equipment or consumables?}
& \rot{Compliant with widely adopted standards?}
& \rot{Interoperable across jurisdictions?}
& \rot{Inexpensive?}
& \rot{Enable items to be located in physical space?}
& \rot{Support additional features or special uses?}
& \rot{Provide high information transfer rates?}
& \rot{Support decentralised control of functionality?}
& \rot{Provide redundancy?}
& \rot{Services are highly available?}
& \rot{Preserve confidentiality?}
& \rot{Ensure data integrity?}
& \rot{Enable transaction audit?}
& \rot{Enforce authentication?}
& \rot{Enforce authorisation?}
& \rot{Enforce access control?}\\ \hline
Flexible tag                    & \T & \T & \T & \T &    &    &    &    & \T & \X & \T & \T &    &    & \T &    & \T &    &    & \T &    &    &    &    &    &    &   \\
Rigid tag                       & \T & \T & \T & \T &    &    &    &    & \T & \T &    & \X &    &    & \X &    & \T &    &    & \T &    &    &    &    &    &    &   \\
Adhesive tag                    & \T & \T & \T & \T &    &    &    &    & \T & \X & \Q & \T &    &    & \T &    & \T &    &    &    &    &    &    &    &    &    &   \\
Tag with RFID                   & \Q & \Q & \T & \T &    & \T & \T &    & \T & \Q & \X & \X & \T &    & \Q & \Q & \T &    &    & \T &    &    &    &    &    &    &   \\
Implantable RFID                & \X & \X &    &    &    & \T & \T &    & \X & \T & \X & \X & \T &    & \X & \Q &    &    &    &    &    &    &    &    &    &    &   \\
Plastic enclosure               &    & \T &    & \X &    &    &    &    & \T & \T & \X & \X &    &    & \T &    & \T &    &    & \T &    &    &    &    &    &    &   \\ \hline
Human-interpretable information & \T & \Q & \X & \T & \T &    & \X & \Q &    &    & \T & \T &    & \Q & \T &    &    & \X &    &    &    &    &    &    &    &    & \X\\
Direct part marking             & \X & \X & \T &    &    &    & \X &    & \T & \T & \X & \Q &    &    &    &    &    &    &    &    &    &    &    &    &    &    & \X\\
Bar code                        & \X & \X & \T & \T &    &    & \T &    & \Q & \Q & \T & \T & \T &    & \T &    &    & \Q &    &    &    &    & \T &    &    &    & \X\\
Passive RFID                    & \X & \X & \T &    &    & \T & \T &    & \X & \X & \X & \Q & \T &    & \T & \Q & \T & \Q &    &    &    &    &    &    &    &    & \T\\
Active RFID                     & \X & \X & \T &    &    & \T & \T &    & \X & \X & \X & \X & \T &    & \X & \T & \T & \T &    &    &    &    &    &    &    &    & \T\\ \hline
Automated locator system        &    &    &    &    & \X & \T & \T &    & \Q & \Q & \X &    &    &    &    &    &    &    &    & \T & \T & \Q & \T & \T & \T & \T & \T\\
GPS locator system              &    &    &    &    &    & \T & \T &    & \T & \Q & \Q & \T & \T & \T & \T & \T &    &    & \T &    & \T &    &    &    &    &    &   \\
IPS locator system              &    &    &    &    &    & \T & \T &    & \T &    & \X & \X &    &    & \X & \T &    &    &    &    & \X &    &    &    &    &    &   \\
Choke point locator system      &    &    &    &    &    & \T & \T &    & \T & \Q & \X & \X &    &    & \T & \T &    &    &    &    & \X &    &    &    &    &    &   \\
Human powered locator system    & \T & \T & \Q & \Q & \T & \X & \X &    & \T & \T & \T & \T &    &    & \T & \T &    &    & \X & \X &    & \X & \X & \X & \X & \X & \X\\
Semi-automated locator system   & \X & \X & \T & \T & \T & \X & \Q &    & \T & \Q & \Q & \X &    &    & \T & \T &    &    &    & \T & \X & \Q & \T & \T & \T & \T &   \\
Beacon location system          & \T &    & \X &    &    &    & \T &    & \X & \Q & \T & \Q &    &    &    & \T &    &    &    &    & \X &    &    &    &    &    &   \\ \hline
Aerial image capture            &    &    &    &    & \T &    &    &    &    &    &    & \X &    &    &    &    &    & \T &    &    &    &    &    &    &    &    &   \\
3D capture                      &    &    &    &    & \T &    &    &    &    &    &    & \X &    &    &    &    &    & \T &    &    &    &    &    &    &    &    &   \\ \hline
Automated speech recognition    &    &    &    &    & \Q & \T & \T &    &    & \X & \X &    &    &    &    &    &    &    &    &    & \T &    &    &    &    &    &   \\
Smart-phone sensors             &    & \T & \T & \T &    & \T & \T &    &    & \X & \T & \T &    &    & \T &    & \T &    &    &    &    &    & \T &    &    &    &   \\
Machine-readable form elements  & \T & \T & \T &    & \Q & \T & \Q &    & \T &    & \T &    &    &    & \T &    &    &    &    &    &    &    &    &    &    &    &   \\
Automated forms                 & \T & \X & \T &    & \T & \T & \Q &    & \Q & \X & \X & \T &    &    & \X &    &    &    &    &    &    & \T & \T & \T & \T & \T & \T\\ \hline
Matching labels                 & \T & \Q & \T & \T & \T &    & \T &    & \T &    & \T & \T &    &    &    &    &    &    &    &    &    &    &    &    &    &    &   \\
Workflow metadata               &    &    & \T & \T &    & \T &    &    &    &    &    &    &    &    &    &    &    &    &    &    &    &    &    &    &    &    &   \\
Automated process logic         &    &    &    &    & \T & \T & \T &    & \Q &    &    &    &    &    &    &    &    &    &    &    &    &    &    & \T & \T &    &   \\ \hline
Open identifier standards       & \Q & \Q & \T & \T & \X & \T &    & \T &    &    &    &    & \T & \T &    &    & \T &    &    &    &    &    &    &    &    &    &   \\
Closed identifier standards     & \Q & \Q & \Q & \Q &    & \T &    & \X &    &    &    &    & \X & \X &    &    &    &    &    &    &    &    &    &    &    &    &   \\ \hline
Local applications              &    &    &    &    &    &    &    &    &    & \T & \T &    &    &    &    &    &    &    & \T &    &    &    &    &    &    &    &   \\
Remote applications             &    &    &    &    &    &    &    &    &    & \X & \X &    &    &    &    &    &    &    &    & \T &    &    &    &    &    &    &   \\
Distributed databases           &    &    &    &    &    &    &    &    & \T & \T &    &    &    &    &    &    &    &    & \T & \T & \T &    & \T &    &    &    &   \\
Opportunistic communication     &    &    &    &    &    &    & \T &    & \T & \T & \T & \T &    &    &    &    &    & \X & \T & \T & \T &    &    &    &    &    &   \\
Mobile ad hoc networks          &    &    &    &    &    &    &    &    & \T & \T & \T & \T & \T &    &    &    &    & \Q & \T & \T & \T &    &    &    &    &    &   \\
Mobile data networks            &    &    &    &    &    &    &    &    &    & \X & \X & \Q & \T &    &    &    &    & \T & \X &    &    &    &    &    &    &    &   \\
Self-managed mobile networks    &    &    &    &    &    &    &    &    &    & \T & \T &    & \T &    &    &    &    & \Q &    &    & \T &    &    &    &    &    &   \\
Self-hosted computing           &    &    &    &    &    &    &    &    &    & \T & \T &    &    &    &    &    &    &    & \Q &    &    & \T &    &    &    &    &   \\
Cloud-hosted computing          &    &    &    &    &    &    &    &    &    & \X & \X & \Q &    &    &    &    &    &    &    & \T & \T & \X &    &    &    &    &   \\ \hline
Digital signatures              &    &    &    &    & \Q &    &    &    &    &    &    &    & \T & \T &    &    &    &    &    &    &    &    & \T &    & \T &    &   \\
Digital certificates            &    &    &    &    & \Q &    &    &    &    &    &    &    & \T & \T &    &    &    &    &    &    &    &    &    &    & \T &    &   \\
Encryption                      &    &    &    &    &    &    &    &    &    &    &    &    & \T & \T &    &    &    &    &    &    &    & \T &    &    &    &    &   \\
Access control                  &    &    &    &    &    &    &    &    &    &    &    &    &    &    &    &    &    &    &    &    &    &    &    & \T & \T & \T & \T\\
Audit trails                    &    &    &    &    &    &    &    &    &    &    &    &    &    &    &    &    &    &    &    &    &    &    &    & \T &    &    &   \\
Block chains                    &    &    &    &    &    &    &    &    &    &    &    &    &    &    &    &    & \T &    & \T & \T & \T &    & \T & \T &    &    &   \\ \hline
\end{tabular}
\endgroup
\vspace{2ex}
    \caption{A static view of the searchable table provided as Supplementary Information indicating how each technology (Table~\ref{tab:technologies}) has ({\T}) or lacks ({\X}) various desirable attributes (Table~\ref{tab:attributes}), or where an attribute may depend on implementation ({\Q}). Technology-attribute combinations that are not meaningful are left blank.}
    \label{tab:tech.attr}
\end{table}

\subsubsection{Digital Domain technologies}
\label{sec:digital}

The primary motive for getting information into the digital domain is to take advantage of information technology (IT), e.g.,  to store, communicate, manipulate, integrate, analyse, search, retrieve, manage and make sense of information. Given our focus on linking physical and digital domains we have organised our presentation under the broad headings of information standards, infrastructure and security. 

\textbf{Information standards}  are fundamental to information exchange within and between systems. Worldwide, we rely on a wide range of standards from organisations like the IEEE and ISO. Most critical to linking physical and digital information in DVI are identifier standards which fall into two categories:
\begin{itemize}
    \item \textbf{Open identifier standards}, such as the Global Standard 1 system, designed to help organisations identify, capture, share and use information (GS1, n.d.). 
    \item \textbf{Closed identifier standards}, local schemes for creating unique digital identifiers specific to events, jurisdictions or organisations. Australian DVI uses identifier standards specific to different states and organisations.
\end{itemize}

\textbf{Information infrastructure} describes the computing, telecommunication and storage systems that support the usage and transmission of information in DVI

Disasters often happen in places or times when communication is limited.  \textbf{Local applications} do not require constant communication, with data stored on the device and communicated later.  \textbf{Remote applications} require constant network access, but are easier to build and maintain.  \textbf{Distributed databases} can provide robust data storage.

When traditional \textbf{Mobile data networks} (used by mobile phones) are not available, \textbf{mobile adhoc networks (MANETS)} or \textbf{self-managed mobile networks} can provide communication to a locally established server.  \textbf{Opportunistic communication}, whereby mobile devices automatically forward data via other devices can help get data back to home base.

Much of the modern internet relies on \textbf{cloud computing} where services are provided from remote data centres to take advantage of economies of scale.  However, \textbf{self-hosting} services on computers that can be taken to the field can provide services where internet access does not exist.

\textbf{Information security} means ensuring that information infrastructure is only used in approved ways.  \textbf{Digital signatures} and \textbf{digital certificates} can be used to ensure the origin and authenticity of data relating to remains.  \textbf{Encryption} prevents unauthorised parties from reading DVI data.  \textbf{Access control} is used to ensure that only authorised parties can read/write data or perform other actions.  Audit trails can be used for quality control of DVI processes.  \textbf{Permissioned block chains} can be used to reliably share information between jurisdictions without giving control to any one party.

We remind the reader that we present a comparative analysis of all candidate technologies in the Supplementary Materials.

\section{Discussion}
Our interviews of Australia DVI practitioners have revealed a range of important issues to consider in designing or deploying technology to improve linkage between physical evidence and digital information. The application of \emph{protocols and training}; the impacts of \emph{stress} and \emph{stressors}; the \emph{plurality of information capture and management systems}; and the \emph{practicalities and constraints} of DVI work each emerged as themes in participant responses and shaped our review of candidate technologies.

While our review presents each candidate individually, they are often interdependent and need to be deployed in combination, e.g., \emph{direct part marking} would likely require a \emph{rigid tag medium} to support some kind of \emph{bar code} conforming to an \emph{identifier standard} that links physical domain labels to an \emph{information infrastructure}. Similarly, technologies for tagging and tracking have to work as part of larger systems, processes and priorities. Technologies need to be considered, understood and improved \emph{in context} i.e., with reference to the broader systems of which they are a part. 

Interdependencies make technological change complex, as does the need to consider the impacts of change on multiple stakeholders. We suspect a consequence of this is for people and teams to develop local practices that enhance tagging and tracking within their local work context, while still fitting in with larger systems. Team-focussed work provides incentives to adapt tools to increase effectiveness and cooperation \cite{orlikowski_using_2000}. People tend to work within their teams, organisations, jurisdictions and languages, becoming familiar with practices that enable and enhance their work within those groups. Often, these practices are widespread (though not universal) e.g., the adoption of INTERPOL protocols. Sometimes, these practices are national (e.g., the use of the NMPVS) or state-level (e.g., the Queensland Police Scene Recovery Booklet). Daily interactions foster practices, customs and cultures that are local to organisations, departments, teams and individuals. 

Practices and challenges within DVI are also context-specific. For instance, the problem of locating remains within a single freezer unit could be solved by simple beacon location systems. However, if those remains are to be linked to additional data (e.g., antemortem or post-mortem) further levels of technological support are required: post-mortem tests may require remains to be transported or that there be auditing to show who had accessed the remains and when; what tests had been performed; and where the remains are. In this scenario, information about the remains would need to be read from and written to a database. While automated or semi-automated systems could be developed (e.g., using RFID), this approach might be best suited to a fixed mortuary facility, rather than a temporary mortuary at a disaster scene. 

There could also come a point when local practices ``speciate'', i.e., become so different that they are no longer compatible with practices elsewhere (for example, in countries where INTERPOL processes are not embraced). This presents real challenges when circumstances demand coherent, coordinated and consistent action from many different teams, as is the case in international disasters like the Boxing Day tsunami or, more frequently, air crash disasters. However, local operations can also provide an environment for experimentation with new technology as well as reinforcement of practice.

Training is clearly a critical pathway to technology adoption. Regular training enables DVI practitioners to adapt to sudden, catastrophic events without the additional mental load of learning new ways of doing things.   Participants said their training relies upon one set of protocols and technologies to be applied in all situations.  Thus any newly introduced technologies or practices should be careful not to negate previous training and be applicable to the wide variety of  DVI contexts.  Naturally, this constrains which technologies can be adopted.

Training also establishes trust in terms of \emph{credibility} (as in ``this process/technology is a credible means to achieve a desired end'') and reliability (as in ``this process/technology will perform predictably well''). Ideally, the benefits of change will be so compelling that desire for adoption will spread by word-of-mouth but realistically, training will also be critical.

This, in turn, leads to the question: what are the opportunities and avenues for training those who would need to use new technology? We note that large-scale simulation events play an important (though infrequent) part of preparing for major disaster and emergency response. These events demand extensive preparation and are probably not the right place to experiment with novel approaches. Small-scale simulation, perhaps even involving virtual reality, seems a more promising avenue, and may also aid in the maintenance of DVI skills between real-world simulation events and actual disasters.

New technologies must work across the physically challenging contexts of actual disasters. As such, designing or choosing technologies to support DVI systems requires an awareness of how these technologies may fail, yet continue to work well enough for the DVI process to be completed. For example, providing labels with identifiers that are both human- and machine-readable and writable allows for the human to take over when the machine is unavailable, and enables the cross-checking of human use when the machine is available. Identifiers exist in a great range of formats and, while care should be taken to not impact human readability, an increased number of identifiers on a single label would ensure its continued usefulness over a greater number of contexts and over time. This approach of ensuring graceful degradation of the transmission of information can be used with a number of DVI technologies. 

Proposed technologies also have to maintain interoperability with associated systems (e.g., police databases) over time. In scientific research, the guiding principles of Findability, Accessibility, Interoperability, and Reusability (FAIR) have been established with this issue in mind \cite{wilkinson_fair_2016}. Upgrading or changing systems often entails a significant financial investment, and this factor is one of the main reasons IT solutions fail to be adopted in emergency management \cite{reddick_information_2011}. Relatedly, adoption at the international level is likely to be a slow and staged process; it is also critical that any suggested technologies remain usable and relevant after a period of time has elapsed. 

For this reason we make a strong recommendation for engaging with tried and true technologies and systems already in use in DVI associated domains. We note that the GS1 information standard has had a distinct focus on healthcare since 2005, encompassing products (e.g., medical devices, pharmaceuticals, vaccines), personnel (e.g., staff, patients) and processes (including sample handling in pathology). However, GS1 is yet to be embraced in DVI operations, at least in Australia where police and pathologists use closed identifier standards, local to each jurisdiction. One participant described their concerns about using an open system:

\begin{quote}
\dots if one of these comes out in transit\dots someone can QR and scan it and get the information off it\dots That worries me. We've just got a straight bar code that will mean nothing to anyone that finds it.---WS
\end{quote}
This caution is understandable, however GS1 supports identifiers that contain a mix of public and private information: a label could reveal only that it relates to a particular entity (e.g., the Australian Federal Police) while keeping more detailed information ``opaque'', i.e., secure and interpretable only by that entity.

\section{Conclusions}
DVI is a complex process performed in extraordinary circumstances by a diverse range of actors from first responders through to specialists in forensics, pathology and disaster response. By linking physical evidence to digital information, how can technology help us tag and track the victims of disaster? We have explored this motivating question by considering technology in the human system of DVI, mindful that it is not just usefulness and usability, but also the context and social practices around technology that affect its successful adoption. 

Our interviews with Australian DVI practitioners elicited four main themes that systematically influence practice and shape candidate technology choice:
\begin{itemize}
    \item \textbf{Protocols, and the training processes that embed them}, are critical to coordinating the efforts of the many people involved in DVI. Training exercises provide a natural pathway to introduce, test and embed technological change.
    \item DVI involves significant \textbf{stress and a range of stressors on practitioners.} Efforts to introduce new technology should be sympathetic to human factors by making key tasks easier to execute; reducing mental load and unnecessary repetition; and either avoiding opportunities for human error or flagging instances of human error.
    \item DVI in Australia uses a \textbf{plurality of information capture and management systems} which have tended to evolve in parallel, rather than in strong coordination, at local, state, national and international levels. We anticipate that this is also the case for DVI in many other countries. Accepting that wholesale global redesign of these systems is unlikely, technology that improves information exchange would be welcomed, provided it enhances data security, business continuity and respects different jurisdictions. 
    \item DVI faces a host of \textbf{practicalities and constraints}, from the physical challenges of the scene, to the social concerns of trust and teamwork under stress, as well as time and resource pressures. These issues affect technology choices both in terms of the extent to which the technology is practical to a situation and also whether a technological solution is the best investment of limited resources.
\end{itemize}

We synthesised insights from study participants with our own expertise and research to produce ten high-level considerations for technology to improve the linkage of physical evidence and digital information in DVI. We then characterised a range of technologies in terms of how their attributes meet these considerations, the details of which are available in searchable form as Supplementary Information.

In abstract terms, this study evaluates different technologies for adoption to help humans (DVI practitioners) link physical evidence with digital information in a particular situation (disaster and emergency). We think that our approach could provide a useful framework for considering candidate technologies in other settings (e.g., healthcare, policing) where a range of actors and systems must cooperate to achieve high-quality outcomes.

Starting with a holistic view of existing systems, processes, practices, preferences and social considerations reminds us that technology has to be accepted and adopted before it can yield its potential benefits.

\section*{Disclosure statement}

No relevant financial or non-financial competing interests to declare.

\section*{Funding}
This research was funded by Queensland University of Technology’s Institute for Future Environments.

Author 1 contributed to the study conception and design, as well as the analysis of data, synthesis of findings, and drafting of the manuscript. Authors 2 and 3 contributed to the study design, and carried out the interviews, data analysis, synthesis of findings, and drafting of the manuscript. Author 4 conceived the study, and participated in its design, synthesis of findings, and drafting of the manuscript. Author 5 contributed to the study conception and design. Author 6 contributed to study conception.  All authors contributed to and approved the final text.


\bibliographystyle{tfnlm}
\bibliography{DVI-FSR-arXiv}

\end{document}